\documentclass[twocolumn,showpacs,preprintnumbers,amsmath,amssymb]{revtex4}

\usepackage{graphicx}
\usepackage{dcolumn}
\usepackage{bm}

\begin{document}


\title{Determination of the freeze-out temperature\\by the isospin thermometer}

\author{P. Napolitani} 
 \email{p.napolitani@gsi.de} 
\author{K.-H. Schmidt} 
\author{P. Armbruster} 
\author{A.S. Botvina}  
 \altaffiliation[Permanent address: ] {Inst. for Nuclear Research, Russian Academy of Sciences, 117312 Moscow, Russia}
\author{M.V. Ricciardi}
\affiliation{GSI, Planckstr. 1, 64291 Darmstadt, Germany\\
}%

\author{L. Tassan-Got, F. Rejmund}
\affiliation{IPN Orsay, IN2P3, 91406 Orsay, France\\ 
}%

\author{T. Enqvist}
\affiliation{University of Jyv\"askyl\"a, 40351 Jyv\"askyl\"a, Finland\\ 
}%

\date{\today}

\begin{abstract}
The high-resolution spectrometer FRS at GSI, Darmstadt, 
provides the full isotopic and 
kinematical identification of fragmentation residues in 
relativistic heavy-ion collisions. Recent measurements 
of the isotopic distribution of heavy projectile fragments 
led to a very surprising new physical finding: the residue 
production does not lose the memory of the $N/Z$ of the 
projectile ending up in a universal de-excitation corridor; 
an ordering of the residues in relation to the neutron 
excess of the projectile has been observed.  
These unexpected features can be interpreted as 
a new manifestation of multifragmentation. 
We have found that at the last 
stage of the reaction the temperature of the big clusters 
subjected to evaporation 
is limited to a universal value. The thermometer to measure 
this limiting temperature is the neutron excess of the residues.
\end{abstract}

\maketitle

\section{\label{sec:level1}Introduction}

Different mechanisms of fragment production can be studied 
within peripheral nucleus-nucleus collisions: Spallation and 
fission have been under investigation for many years. In the last decade 
a so-called multifragmentation reaction, which leads to the total 
disintegration of heavy nuclei into light and intermediate-mass 
fragments (IMF) arouse large interest 
\cite{ALADIN_MSU,INDRA,EOS}. Originally it was motivated 
by studying the liquid-gas type phase transition in nuclear matter 
\cite{Bondorf95,Pochodzalla95}, as well as the role of thermal and spinodal 
instabilities in the disintegration of finite nuclei. 
At high-energy collisions the multifragmentation share was found to correspond to 
10--20 $\%$ of the total reaction cross-section, and its contribution 
to the yield of the IMF (with charges $Z$=3--30) is crucially 
important. With increasing excitation energy of the thermal source, the 
transition from the evaporation and fission decay mechanisms 
to the multifragmentation is smooth: The probability for the formation
of one compound nucleus 
decreases, whereas the multifragmentation appears first as a freeze-out 
state with two hot fragments and progressively involves three, four, and 
many fragment channels with increasing excitation. 
In this mechanism, a considerable part of the available excitation 
energy goes into the disintegration of the system, but not into increasing 
the temperature of the fragments \cite{Bondorf95}. 

\section{Experimental investigation of the reaction mechanism}
\begin{figure*}
\includegraphics{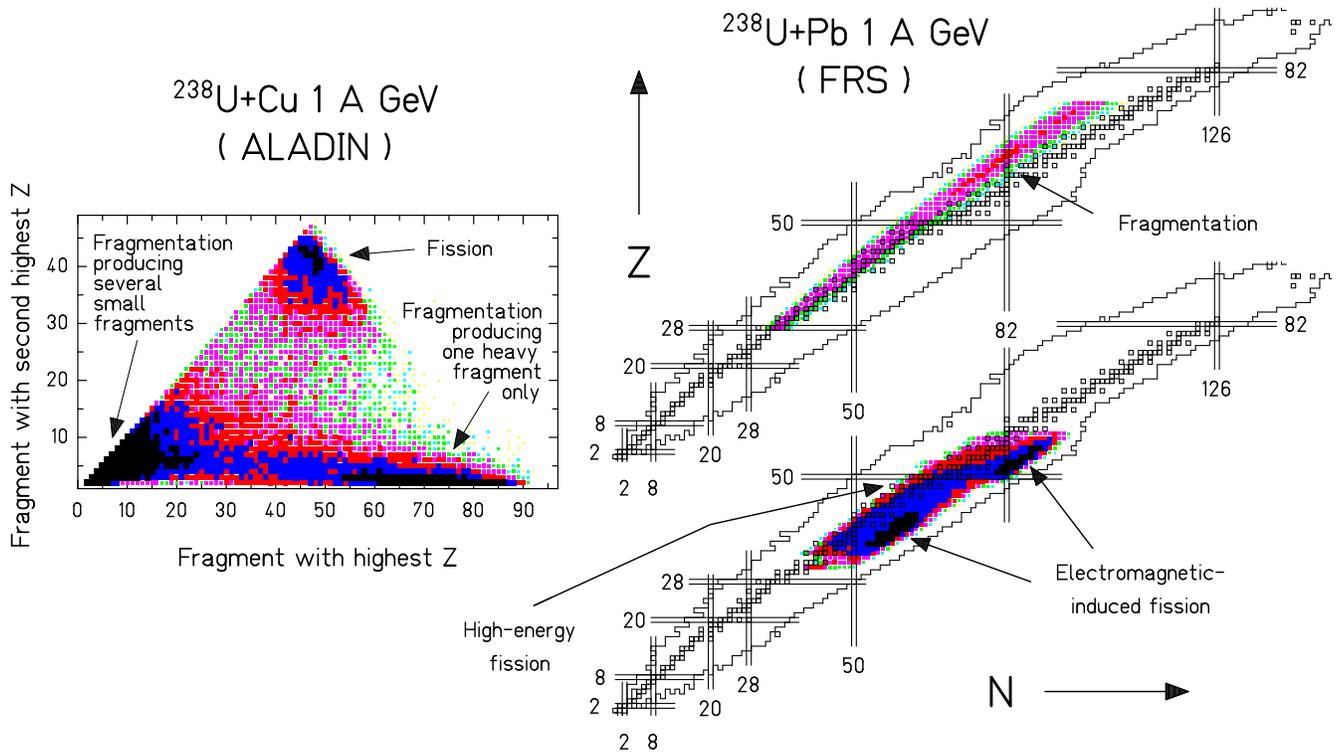}
\caption{\label{fig:wide}
Comparison between the experimental investigations that could be 
performed with ALADIN and with FRS. Left. ALADIN provides the 
measurement of the multiplicity and the correlation between the 
emitted particles for each event. Right: FRS provides the full 
isotopic and kinematical identification of one fragment for each 
event: it is possible to reconstruct the reaction process.}
\end{figure*}

Multifragmentation is the field of intense investigation 
of the ALADIN (GSI, Darmstadt) collaboration 
\cite{ALADIN_MSU}, with an instrument whose total 
acceptance allows for the full counting of the produced 
fragments and their correlations. This precious information 
could be analyzed as in figure 1 (left), where a study of 
{\bf $^{238}$U} impinging on a copper target 
at $1$ $A$ GeV \cite{Schuttauf96,ALADIN_MSU}
is shown: for each collision we have a collection of fragments 
of different $Z$, we chose the two residues having the highest $Z$ 
and plot one against the other for each event. 
The fission products, individuated by two fragments of about 
half the projectile charge, are well separated. The region of 
multifragmentation, characterized by high multiplicity, small 
impact parameter and identified in the low-$Z$ corner shows a 
gradual transition towards more and more peripheral collisions, 
where fragments of about half the projectile charge are correlated 
to very light residues like lithium, beryllium or boron. 
The aim of this paper is to 
discuss the possibility that the heavy fragments could provide a 
complementary information on the multifragmentation process. 

By concentrating on the heavy residue, we can take advantage 
of the high-resolution spectrometer FRS (GSI, Darmstadt) \cite{FRS},
which was 
designed to obtain exclusive information on the heavy fragments. 
The precise kinematical identification of the FRS provides 
unambiguous information on the de-excitation process that 
generated the observed residues: the measurement of the velocity 
distribution of each fragment enables to clearly disentangle 
fragmentation and fission products. In addition, the measurements
of the energy loss and 
the mass-to-charge ratio lead to the full isotopic 
identification. A very systematic overview is shown in picture 1 (right) 
for the collision of {\bf $^{238}$U} with lead at $1$ $A$ GeV 
\cite{Enqvist99}: we can clearly 
recognize the region of electromagnetic-induced fission: this process 
originates from low-excited projectile-like prefragments and, 
consequently, produces very neutron-rich residues. The area of 
high-energy fission, generated by more excited prefragments, is 
less neutron rich. Fragmentation products, expected to originate 
from a long evaporation process starting from very highly excited 
prefragments, populate the neutron-deficient side of the 
isotopic chart.           
This analysis, based on the knowledge of the reaction process,
establishes a connection between the isotopic distribution of the
residues and the excitation energy introduced at the beginning of
the evaporation. In the following, we will restrict this analysis
to fragmentation only and continue to study how the neutron
excess can provide indications on the reaction mechanism.

\section{Study of the neutron excess of the fragmentation residues}
\begin{figure*}
\includegraphics{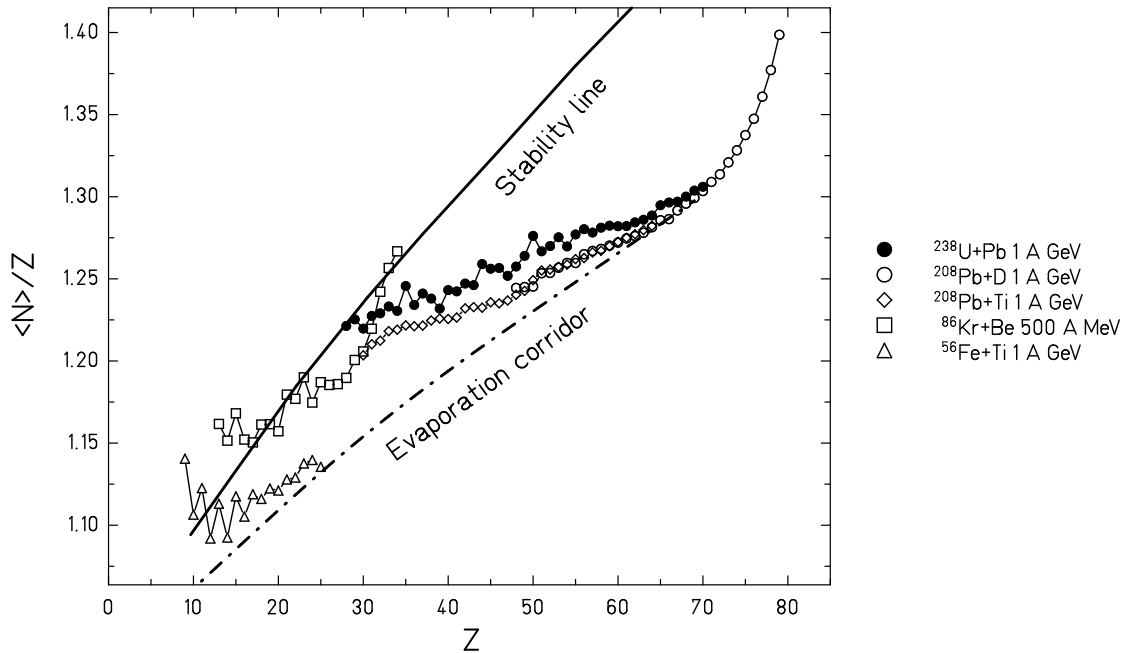}
\caption{\label{fig:wide} 
Collection of experimental data showing a deviation from 
the evaporation corridor. The projectiles have different 
neutron excess. {\bf $^{56}$Fe} \cite{Napolitani01} is not neutron rich and its 
residues approach the evaporation corridor. {\bf $^{86}$Kr} \cite{Weber92} does 
not reach the corridor. The lead system, {\bf $^{208}$Pb} on deuterons \cite{Enqvist01} 
and {\bf $^{208}$Pb} on {\bf Ti} \cite{Enqvist01b}, shows a clear deviation from the corridor. 
The {\bf $^{238}$U} system \cite{Enqvist99} deviates and is more neutron rich than {\bf $^{208}$Pb}. 
The evaporation corridor has been reproduced with the two-stage model 
(the abrasion code ABRA coupled with the evaporation code ABLA).
}
\end{figure*}

The classical model describes fragmentation as a two-stage process \cite{Gaimard91}: 
highly excited prefragments are generated in an initial fast stage, 
usually described as an abrasion process (e.g. for 
nucleus-nucleus collisions) or an intra-nuclear cascade 
process (mostly for hadron-nucleus collisions). 
The time required by the fast stage is of the order 
of $20$ to $50$ fm/c. The resulting prefragment, whose neutron-to-proton 
ratio is initially close to the projectile, 
will then experience a sequential decay (or evaporation) 
dominated by neutron emission. The role of the second process 
is to change the neutron-to-proton ratio and to guide 
the isotopic production towards a universal evaporation corridor, 
where the initially more favourite neutron emission is finally 
balanced by proton emission. If the excitation energy introduced 
by the fast stage were entirely removed by sequential decay, 
the evaporation would be long enough for the residues to end up 
in the universal corridor and lose the memory of the projectile.

This classical picture is in contradiction with recent 
experimental data on fragmentation of neutron-rich projectiles. 
The isotopic cross-sections of {\bf $^{238}$U} fragments, produced in a lead 
target at $1$ $A$ GeV \cite{Enqvist99}, show an increasing deviation from the 
corridor (represented by the two-stage model calculation \cite{Gaimard91} 
in figure 2) for decreasing Z towards higher neutron numbers. 
As presented in figure 2, the measured fragmentation residues 
of {\bf $^{238}$U} even have the tendency to cross the stability line; 
this is very surprising because, with respect to $\beta$ stability 
(equal Fermi levels of protons and neutrons), the evaporation 
of protons is suppressed by the Coulomb barrier. 

The increasing deviation from the evaporation corridor indicates 
that, for collisions expected to be more and more violent, less 
and less excitation energy is available for the evaporation stage; 
evidently, in the classic picture of the fragmentation mechanism, 
an intermediate process that removed this excess of energy is 
missing. 

Figure 2 presents the evolution of the mean neutron-to-proton 
ratio of the residues produced by the fragmentation of different 
systems. We infer that there is an ordering of the residues in 
relation to the neutron excess of the projectile: the isotopic 
distributions of different fragmenting systems do not collapse on 
the same universal evaporation corridor, but they are more neutron 
rich for more neutron-rich projectiles, showing an evident memory 
effect related to the neutron-to-proton ratio of the 
projectile.

\section{A thermometer based on the neutron excess}

We can infer that the present data (figure 2) are 
closely related to the observation of multifragmentation. We will
analyze the data with two models. First, a three-stage model
is used, which describes the multifragmentation as an
intermediate break-up stage after
the high-energy nucleon-nucleon collisions and before the
sequential decay in a rather schematic way. Secondly, we have
chosen the Statistical Multifragmentation Model (SMM) as a dedicated 
model for multifragmentation, which is very effective in the description 
of the experimental data (see e.g. \cite{Bondorf95,Botvina95}). 
The most intensively investigated signature for the onset of 
multifragmentation is the production of several about equal-size fragments.
However, also in accordance with SMM, the fragments do not need to be
necessarily about equal-size.
In peripheral collisions, we could 
expect a disassembly of the hot primary fragment into a heavy 
residue, accompanied by clusters and very light nuclei: in this 
case, when one heavy fragment is observed in the FRS, we still have
indication of the onset of the multifragmentation phenomenon 
\cite{Bondorf95}. 
\begin{figure}[!b]
\includegraphics[width=1\columnwidth]{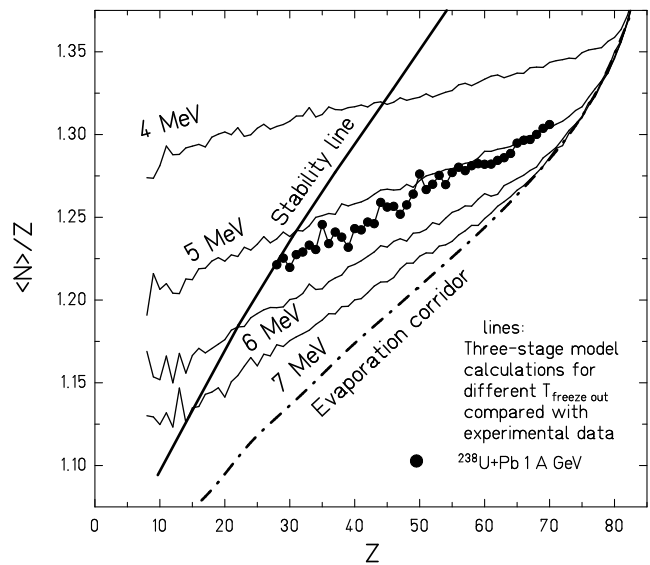}
\caption{\label{fig:epsart} 
The principle of the 'isospin thermometer'. 
The value of the freeze-out temperature is changed as a 
parameter of a three-stage model \cite{Gaimard91} calculation 
(ABRA followed by a break-up simulation and later by ABLA). 
The experimental data of fragmentation of {\bf $^{238}$U} are reproduced 
for a freeze-out temperature between $5$ and $6$ MeV.
}
\end{figure}

The experimental data can be reproduced introducing an intermediate 
process that, right after the fast stage, removes part of the mass 
and energy. This stage can be described as a break-up process: the 
compression caused by the high excitation energy and, eventually, 
by the collision dynamics provides a high internal pressure. 
A consequent expansion and disassembly of the system will remove 
part of the initial excitation energy. This process might be
related to a liquid-gas-like phase transition. 
A fundamental assumption for the process is the 
conservation of the mean neutron to charge-number ratio $<N>/Z$. 
The nearly conservation of the $<N>/Z$ ratio is also predicted by 
SMM \cite{Botvina01}. The break-up stage 
ends when the system reaches the freeze-out transition, and 
re-condensates in an ensemble of cooled fragments. If we assume that 
thermodynamic equilibrium is established in the system at the 
transition point, we can consider a freeze-out temperature as a 
major parameter of the reaction process. This parameter can in fact 
be defined as a limiting temperature, since no more than the 
corresponding excitation energy will be available for the sequential 
decay. As a result, also the length of the evaporation process will 
be limited and determined by this value. We assume that the 
sequential decay will start from an excitation energy corresponding 
to the freeze-out temperature.
\begin{figure}[!t]
\includegraphics[width=1\columnwidth]{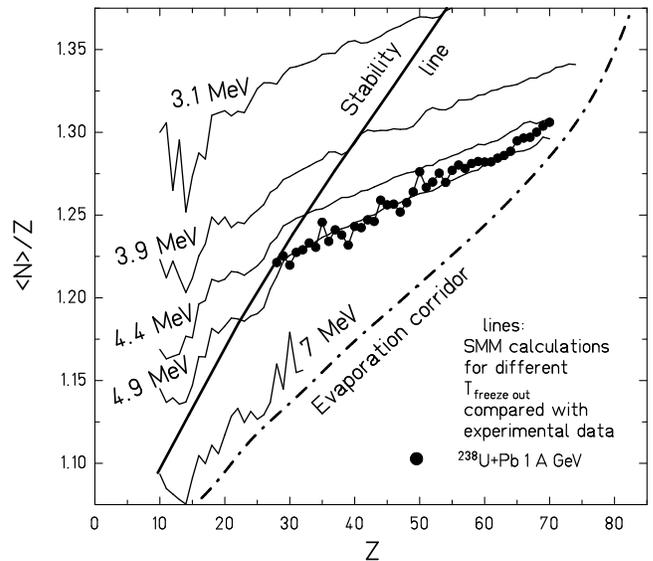}
\caption{\label{fig:epsart} 
Set of calculations performed with SMM\cite{Bondorf95} for different values 
of the excitation energy at break-up corresponding to mean temperatures of
$3.1$, $3.9$, $4.4$, $4.9$ and $7$ MeV .
}
\end{figure}

The deviation of the experimental data from the universal corridor 
suggests a new tool to measure the length of the sequential decay 
and, consequently, the value of the freeze-out temperature. 
Between the fast stage and the evaporation stage a new step has been added 
in order to describe the effect of the break-up. We should  
observe that the treatment of the partitioning of the spectator 
into the break-up products affects the final cross section, but has 
no effect on the neutron excess of the isotopic distribution of 
the residues; the quantity $<N>/Z$ of the residues is only sensible 
to the value of the freeze-out temperature. 
This assumption motivates the calculation shown in figure 3, where the 
experimental data of fragmentation of {\bf $^{238}$U} on a lead target are 
compared with a set of calculations. 
The treatment of the partitioning was simplified by a parameterisation; the
description of the neutron excess of the residues is determined by the
freeze-out temperature as the only free parameter of the calculation:
when this parameter has low values, the 
break-up is dominating and the residues are too neutron rich. 
Inversely, for high values, the break-up tends to be suppressed 
and the isotopic production ends up in the evaporation corridor.

The measurement fixes the freeze-out temperature in a 
range between $5$ and $6$ MeV. 
Missing the complete data for systems 
covering a wide range of the neutron excess, we still cannot 
determine whether the freeze-out temperature is a constant value 
or a function of $N/Z$. However, the choice of around $5$ MeV provides a 
very satisfactory reproduction of the available data \cite{Schmidt02}. 

A more elaborate physical description of the partitioning is provided by SMM.
In Fig.~4 we present the results of a series of SMM calculations for 
the disintegration of a $^{238}$U source. 
The excitation energy at break-up was taken as a parameter. The calculation 
was performed for excitation energies of $1$, $1.5$, $2$, $2.5$ and $8$ MeV,
corresponding to mean temperatures of $3.1$, $3.9$, $4.4$, $4.9$ and $7$ MeV, 
respectively. 
With increasing temperature the break up generates more excited fragments,
and the mean neutron excess of the residues 
approaches the evaporation corridor.
This calculation led to about the same results as presented in Fig.~3: also in this
case a remarkable agreement with the data is obtained for a freeze-out 
temperature of around $5$ MeV. 
An important finding is the independence of the temperature from 
the mass of the residues. 
The observed universality is an indication that the limiting freeze-out 
temperature is rather independent of the initial conditions. 

\section{Effect of the break-up on the mass-distribution}
\begin{figure}[!b]
\includegraphics[width=1\columnwidth]{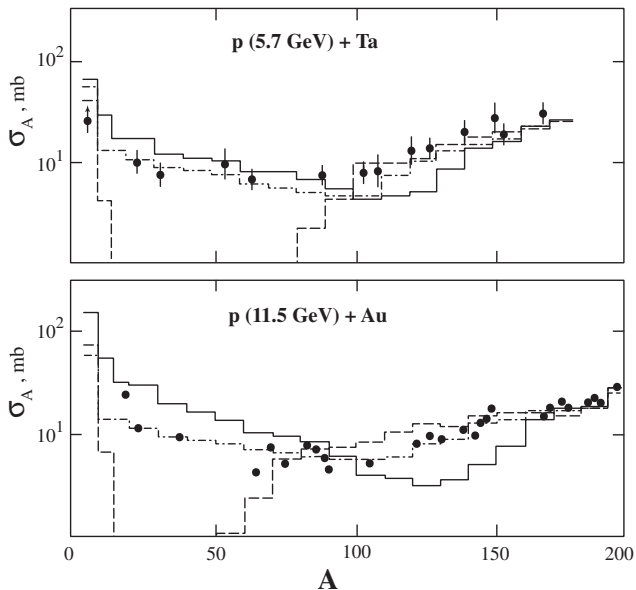}
\caption{\label{fig:epsart} 
Mass yields of fragments in the reaction shown in the figure.
The dots denote the experimental data for tantalum \cite{Grover62} 
and for gold \cite{Kaufman76}.
The histograms correspond to calculations. Dashed lines: cascade-evaporation
model (break-up was disregarded). Solid line: calculation where the break-up is
included. Mashed line: calculation with break-up included, with the addition  of
a preequilibrium process. 
}
\end{figure}
\begin{figure*}
\includegraphics{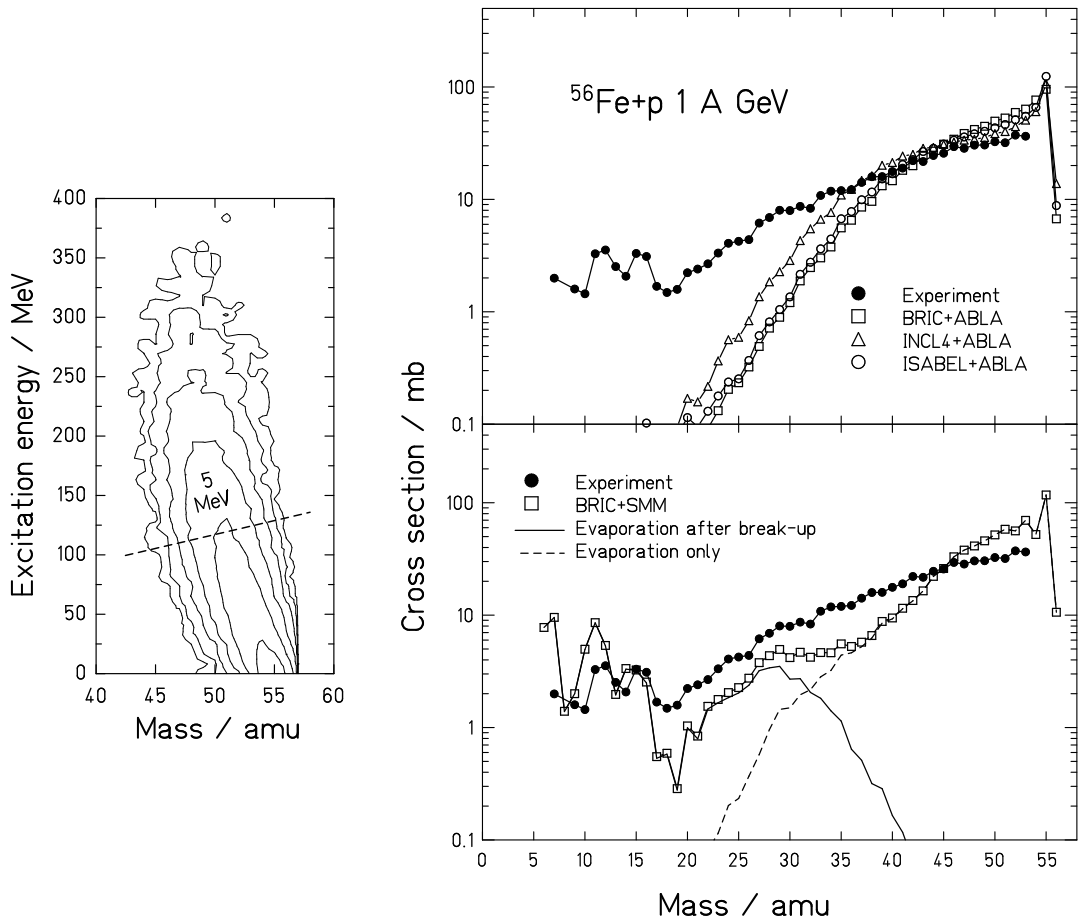}
\caption{\label{fig:wide}
Left. Prefragments produced right after the intra-nuclear cascade stage in the 
reaction {\bf $^{56}$Fe} on proton at $1$ $A$ GeV (calculation performed with 
INCL4 \cite{Boudard01}). The dashed line indicates the excitation energy 
corresponding to a freeze-out temperature of $5$ MeV. Right, top. 
Three intranuclear-cascade codes (BRIC \cite{Duarte}, INCL4 \cite{Boudard01}, 
ISABEL \cite{Yariv79}) coupled with the evaporation code ABLA do not 
reproduce the experimental data of {\bf $^{56}$Fe} on proton 
at $1$ $A$ GeV \cite{Napolitani01}. Right, bottom. The experimental data are 
reproduced if the break-up is considered. The calculation is 
performed coupling the intranuclear-cascade code BRIC with SMM.
}
\end{figure*}

We investigate now the effect of the break-up in recent data of fragmentation
of {\bf $^{56}$Fe} \cite{Napolitani01} in a proton target at $1$ $A$ GeV,
measured in inverse kinematics at the FRS.
This is an interesting case to study the generality of the break-up process
and its extension to cases where the excitation energy 
of the system is neither high enough to be dominated by thermal 
instability, nor too low to exclude some break-up events. 
However, {\bf $^{56}$Fe} is not neutron rich and the isotopic distribution is 
not expected to deviate from the evaporation corridor. To observe 
the effect of thermal instability, we should then look for a 
different signature like, for instance, the mass distribution.

As was established in previous studies (see e.g. references in 
\cite{Bondorf95}), in reactions where a considerable energy is 
transferred to the 
thermal source, the multifragmentation influences essentially the yield of the
residues. The ALADIN fragmentation data provide a typical example of 
multifragmentation obtained in peripheral nucleus-nucleus collisions at high energy 
\cite{Botvina95}. It was also found that this process takes place in 
reactions induced by high-energy protons. In Figure~5, we 
present the results of the analysis performed in \cite{Botvina90} 
concerning inclusive yields of fragments from tantalum \cite{Grover62} 
and gold \cite{Kaufman76} targets. 
The theoretical calculations were done within cascade, evaporation, 
and fission models which ignore the multifragment break-up stage, and, 
alternatively, with the same models including the multifragmentation additionally. 
In the last case the INC (Dubna version) and SMM were used. We can clearly 
see that fragments with $A=10-60$ can be explained only by 
multifragmentation. However, the yield is not fully reproduced with the 
INC+SMM calculations. An additional correction of the parameters 
(masses and excitation 
energies) of the after-cascade residual nuclei is necessary. We can 
attribute this additional preequilibrium process to the expansion of 
the residues toward the freeze-out volume \cite{Bondorf95}. 

We applied the same investigation to study the reaction of
{\bf $^{56}$Fe} on proton at $1$ $A$ GeV.
Such a system is expected to produce
a broad distribution of the residues in excitation 
energy depending on the impact parameter and fluctuations during 
the preequilibrium process.
In general, nucleus-proton reactions at 
about $1$ $A$ GeV generate too low excited system to undergo a 
multifragmentation event but, nevertheless, {\bf $^{56}$Fe} is light enough 
to still reach the break-up temperature. A very indicative way to 
discriminate the possible break-up events is presented in figure 6 
(left): the plot shows a collection of calculated prefragments, 
generated by the fast stage, distributed according to their mass 
and excitation energy. The region above the energy corresponding 
to a freeze-out temperature of $5$ MeV is hot enough to undergo a 
break-up stage. The region below collects the prefragments expected 
to start the sequential decay immediately after the fast stage. 

As shown in figure 6 (top), three intranuclear cascade codes 
(BRIC \cite{Duarte}, INCL4 \cite{Boudard01}, ISABEL \cite{Yariv79}) 
coupled with the same evaporation 
code (ABLA) reveal to severely under-predict the experimental data 
for the lighter half of the mass distribution (figure 6, top). 
Inversely, the introduction of a break-up stage (calculation 
performed with BRIC coupled with SMM) provides a satisfactory 
reproduction of the measurement. 
A slight discrepancy in the 
yields of fragments with mass numbers $A>A_{target}/2$ is probably 
caused by the discussed uncertainty of the parameters of the excited 
residues produced after the intra-nuclear cascade stage: the hybrid
model used to reproduce the data is still too simple to correctly
describe the transition from the fast stage to the break-up, and the 
inclusion of a preequilibrium process could be needed.
In figure 6 (bottom) it is possible 
to observe that the light part of the mass distribution is mainly 
populated by break-up events. 
A very interesting result revealed by 
the experiment is the tendency for the cross sections to increase 
in the region of light masses: the only mechanism able to reproduce 
this characteristic seems to be a series of break-up events.

\section{Conclusions }

The analysis of recent experimental data revealed the influence of
multifragmentation to be more general than expected.
In nucleus-nucleus reactions, the signature for the onset of 
multifragmentation is not only carried by the lightest fragments,
but it extends towards the intermediate-mass fragments.
The break-up process does not only describe the reactions at
small impact parameter,
but it should be taken into account in peripheral collisions
as well.
Moreover, the memory of the $N/Z$ of the projectile reflected in the 
neutron excess of the residues is not only an experimental evidence of 
the generality of multifragmentation, but it is also a new tool to study 
the reaction mechanism.
The study of the relation between the isotopic distribution of the 
residues and the break-up process opens up new possibilities of
investigation: the combination between the isotopic identification
(FRS) and the measurement of the multiplicity (ALADIN) can provide a new
insight about the role of the impact parameter on multifragmentation.

In the case of nucleon-nucleus reactions, the impact of thermal instability
is also more general then expected. 
We have experimental evidence that the fragmentation of light nuclei like 
{\bf $^{56}$Fe} in a proton target at $1$ $A$ GeV, shows similar features 
as in the case of proton-induced fragmentation of heavy nuclei at high
energy (5--10 GeV range).
Disregarding the break-up process in the complete description of
nucleon-nucleus collisions could lead to an underprediction of the 
yields of the light residues by several orders of magnitude, 
even at beam energies around $1$ $A$ GeV.

\end{document}